# Quantum Graph States: Bridging Classical Theory and Quantum Innovation—Workshop Summary

**Eric Chitambar[1], Kenneth Goodenough[2], Otfried Gühne[3], Rose McCarty[4], Simon Perdrix[5], Vito Scarola*,[6], Shuo Sun[7], and Quntao Zhang[8,9]**

[1] Department of Electrical and Computer Engineering, University of Illinois Urbana-Champaign, Urbana, IL, USA
[2] College of Information and Computer Science, University of Massachusetts Amherst, Amherst, MA, USA
[3] Naturwissenschaftlich-Technische Fakultät, Universität Siegen, Siegen, Germany
[4] School of Mathematics and School of Computer Science, Georgia Institute of Technology, Atlanta, GA USA
[5] Inria Mocqua, LORIA, CNRS, Université de Lorraine, Nancy, France
[6] Department of Physics, Virginia Tech, Blacksburg, VA USA
[7] JILA and Department of Physics, University of Colorado Boulder, CO, USA
[8] Ming Hsieh Department of Electrical and Computer Engineering, University of Southern California, Los Angeles, 90089, CA, USA
[9] Department of Physics and Astronomy, University of Southern California, University of Southern California, Los Angeles, 90089, CA, USA
* Contact person and workshop organizer, email: scarola@vt.edu

## Abstract

This workshop brought together experts in classical graph theory and quantum information science to explore the intersection of these fields, with a focus on quantum graph states and their applications in computing, networking, and sensing. The sessions highlighted the foundational role of graph-theoretic structures—such as rank-width, vertex-minors, and hypergraphs—in enabling measurement-based quantum computation, fault-tolerant architectures, and distributed quantum sensing. Key challenges identified include the need for scalable entanglement generation, robust benchmarking methods, and deeper theoretical understanding of generalized graph states. The workshop concluded with targeted research recommendations, emphasizing interdisciplinary collaboration to address open problems in entanglement structure, simulation complexity, and experimental realization across diverse quantum platforms.

## Table of Contents

## I. Workshop Format

The workshop took place May 28th-May 30th, 2025, in Virginia Tech's Executive Briefing Center in Arlington, Virginia, USA. The participant list can be found in the Appendix. The workshop was organized by Vito Scarola and was sponsored by the Air Force Office of Scientific Research, the Office of Naval Research, the Army Research Office, and Virginia Tech's College of Science and Department of Physics.

The first day hosted two tutorials, one on classical graph theory and one on quantum graph states. The following two days consisted of four presentation sessions. Each session was followed by a discussion moderated by an expert in each field. Workshop participants participated in the discussion. Results from these presentations and discussions were recorded and form the basis of the material presented here.

Tutorials

- Vito Scarola: Introduction
- Rose McCarty: Graph Theory Tutorial
- Eric Chitambar: Graph States Tutorial

Session 1: Quantum Graph States and Computing, Chair: Simon Perdrix

- Hans Briegel: Perspectives on Random Graph States and Measurement-based Quantum Computation
- Akimasa Miyake: Graph States for Quantum Computation and Network: From Universality to Quantum Advantage
- Ken Brown: Fault Tolerance and Cluster States

Discussion Moderator: Simon Perdrix

Session 2: Quantum Graph State Experiments, Chair: Shuo Sun

- Olivier Pfister: Qubits Without Qubits: Continuous-variable Measurement-based Photonic Quantum Computing
- David Weiss: Creating Discrete Variable Cluster States with Neutral Atoms, Ions or Molecules: an Overview
- Monika Schleier-Smith: Atoms Interlinked by Light: Nonlocal Graph States for Quantum Sensing & Simulation

Discussion Moderator: Shuo Sun

Session 3: Graphs and Networking, Chair: Kenneth Goodenough

- Maria Chudnovsky: How to Bound Treewidth: Structure and Algorithms
- Asaf Ferber: Quantum Algorithms on Graphs
- Alexey Gorshkov: Graphs and Graph States in Quantum

Discussion Moderator: Kenneth Goodenough

Session 4: Utility of Entanglement in Graphs and Networks, Chair: Mark Wilde

- Otfried Gühne: Some Results on Graph and Hypergraph States
- Quntao Zhuang: Utility of Entanglement in Quantum Sensing (and Transduction)

Discussion Moderator: Mark Wilde

## II. Introduction and Overview

Graph theory provides important and basic tools that overlap with work in quantum information science, particularly in computing, error correction, networking, and sensing. The goal of the workshop was to 1) Promote the connection between mathematical graph theory and distributed quantum information processing, emphasizing concepts relevant for multiple application areas including quantum computing, communications, sensing, etc. and 2) Bring together classical graph theorists and quantum information scientists to gauge the most interesting basic research challenges relevant for both disciplines.

This section provides overviews on graph theory and quantum graph states based on workshop tutorials from Rose McCarty and Eric Chitambar, respectively.

*Classical Graph Theory and Its Modern Frontiers*

Graph theory, a foundational area of discrete mathematics, provides a versatile framework for modeling relationships and interactions in complex systems. From transportation and communication networks to quantum computing, graphs offer a powerful abstraction for analyzing connectivity, flow, and structure. This workshop began with a tutorial by Rose McCarty, who presented a comprehensive overview of classical graph theory and its modern applications, bridging foundational concepts with emerging trends in quantum information science. At its core, a graph consists of a set of vertices (nodes) and a set of edges (connections between nodes), see Figure 1. This simple structure enables the modeling of diverse systems. For instance, a road network can be represented as a graph where intersections are vertices and roads are edges. By assigning weights to edges—such as travel time or capacity—graph theory facilitates the analysis of optimal paths and network efficiency.

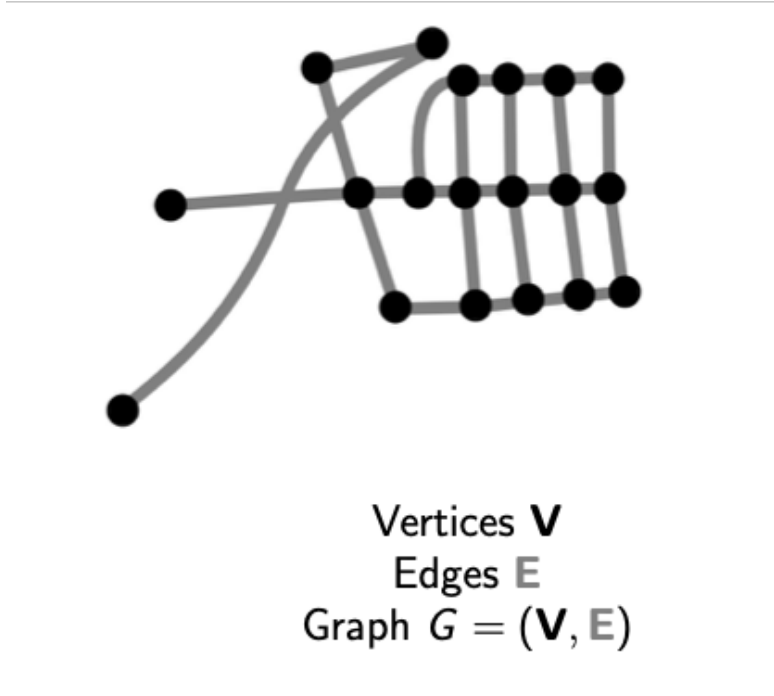

*Figure 1: A generic graph*

One of the most celebrated results in classical graph theory is the Max-Flow Min-Cut Theorem [FoFu56]. This theorem states that the maximum flow from a source to a target in a network equals the capacity of the smallest set of edges whose removal disconnects the source from the target. This principle underpins applications in transportation, electrical grids, and data routing. Examples include determining how much electricity can be sent from one node to another and identifying vulnerabilities in power grids.

The concept of graph expansion is central to understanding network robustness [HoLW06]. A graph is a poor expander if removing a few vertices or edges can disconnect a large portion of the graph. Good expanders, by contrast, are resilient and support efficient communication and error correction. Cliques are ideal expanders, however it is often desirable to construct expanders with fewer edges. These sparse expanders can be constructed using probabilistic methods or algebraic techniques.

Modern graph theory explores high-dimensional analogs of graphs, such as 2D and 3D expanders [Lubo19]. These structures include not only edges but also higher-order connections like triangles and tetrahedra. A 2D expander, for example, requires that both the graph and the neighborhood of each vertex (projected onto triangles) exhibit expansion properties. This area is particularly relevant to error correction, pseudorandomness, and quantum computing.

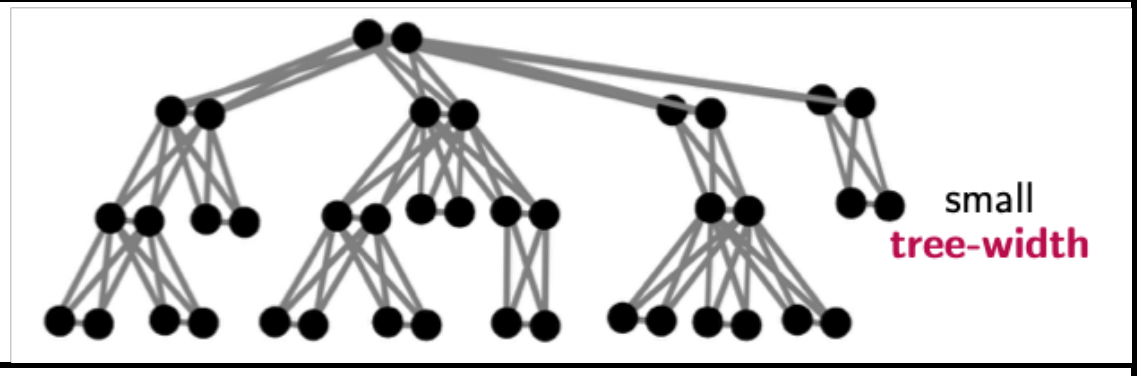

*Figure 2: A graph demonstrating small tree width.*

Structural properties in graphs are important.  Planar graphs, semi-algebraic graphs (e.g., radar coverage networks), and graphs with small tree-width (Figure 2) or rank-width exhibit special characteristics that can be exploited algorithmically. For example, semi-algebraic graphs arise when radar towers are connected to locations they cover, forming edges based on polynomial inequalities. These structures often appear in geometry, biology, and network science.

A particularly exciting frontier is the intersection of graph theory and quantum computing. In the measurement-based model of quantum computation, graph states—quantum states associated with graphs—serve as computational resources [ChGo19]. The entanglement structure of these states, captured by the graph's topology, determines the computational power of the system. Local Clifford (LC) equivalence and local complementation (discussed below) provide a graph-theoretic lens for understanding quantum state equivalence.

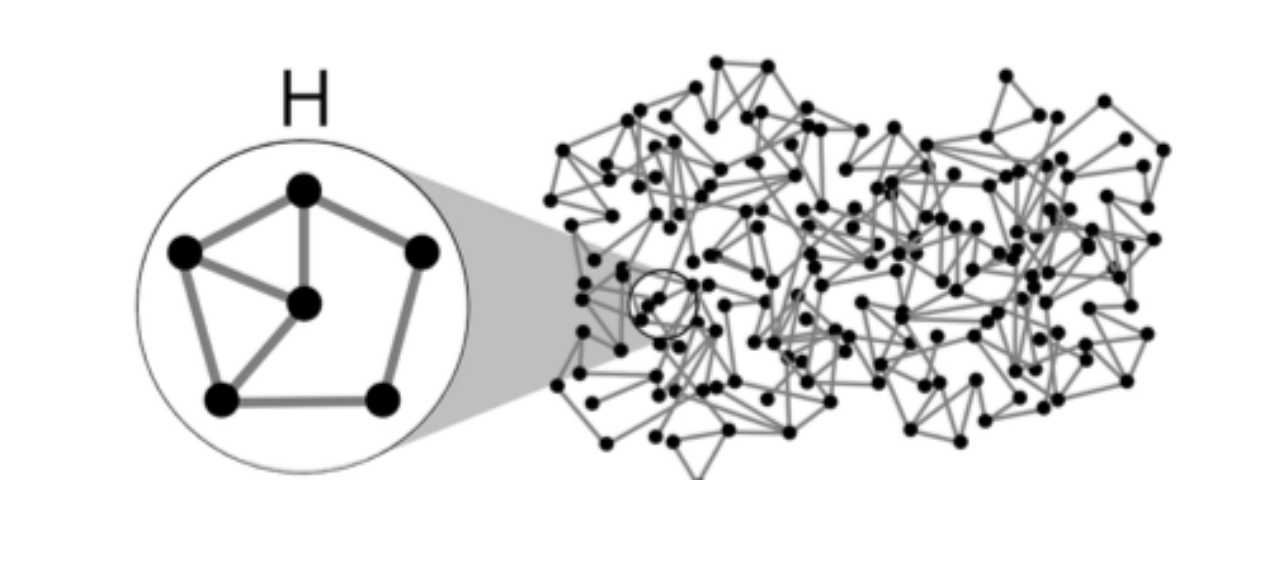

*Figure 3: A graph G forbids a graph H as a vertex-minor if no graph that is a local complement to G contains a copy of H.*

A central conjecture by Geelen [Mcca21] posits that if a graph forbids a fixed vertex-minor (Figure 3), then quantum computation on the corresponding graph state can be efficiently simulated classically. This conjecture connects structural graph theory with quantum complexity and has implications for identifying the minimal resources required for quantum advantage. It suggests that connected subgraphs of a quantum resource state without vertex minors cannot be used to implement computations that are more powerful than classical computing, emphasizing the importance of graph structure in quantum algorithms.

*Quantum Graph States and Measurement-based Quantum Computing*

In recent years, graph theory has found a profound new role in quantum computing, particularly through the study of quantum graph states [HDER06]. A graph state is a specific type of quantum state that is defined by a graph: each vertex corresponds to a qubit initialized in the $|+\rangle$ state, and each edge represents a controlled-$Z$ (CZ) entangling operation between the connected qubits. These states are a subset of stabilizer states, which are quantum states uniquely defined as the simultaneous +1 eigenstates of a set of commuting Pauli operators.

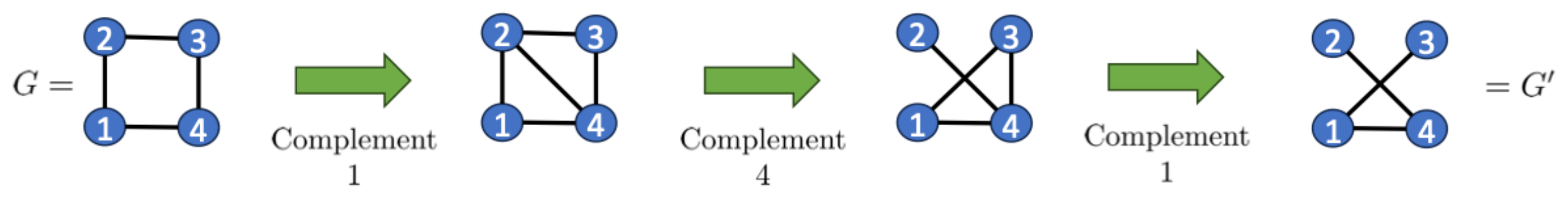

$$|G'\rangle = \left(-e^{i\frac{\pi}{2}\frac{X_1 - Z_1}{\sqrt{2}}} \otimes Z_2 \otimes Z_3 \otimes e^{i\frac{\pi}{2}\frac{X_4 - Z_4}{\sqrt{2}}}\right)|G\rangle.$$

*Figure 4: Examples of local complementation between graph states* G *and* G'

The stabilizer formalism provides a compact and efficient way to describe large, entangled quantum states and certain types of quantum algorithms. In this framework, the stabilizer group of a graph state is generated by operators of the form $X_i \prod_{k\in\mathcal{N}(i)} Z_k$, where the Pauli $X_i$ operator acts on a vertex $i$ and the Pauli $Z_k$ operator acts on vertices $k$ in the neighborhood of $i$, $\mathcal{N}(i)$. This structure allows for a binary vector representation of the stabilizers, enabling the use of symplectic geometry and linear algebra to analyze and manipulate quantum states.

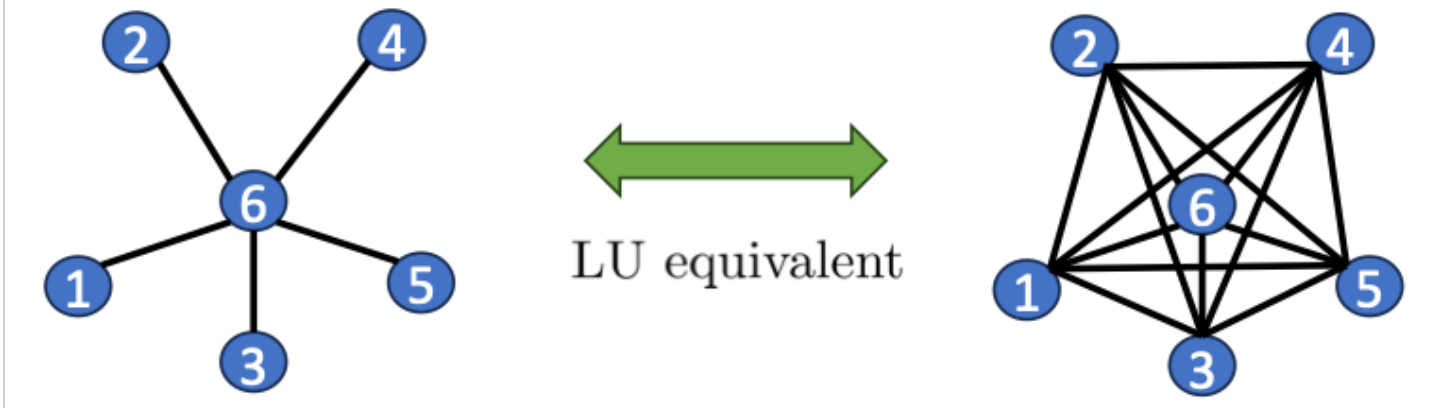

*Figure 5: Example of LU equivalent graph states*

A central concept in the study of graph states is local Clifford (LC) equivalence [VaDD04, HDER06]. Two graph states are LC-equivalent if they can be transformed into one another using only single-qubit Clifford operations. This equivalence can be characterized graph-theoretically through local complementation, an operation that modifies the neighborhood of a vertex by complementing the subgraph induced by its neighbors (Figure 4). LC-equivalence provides a powerful tool for classifying entanglement and understanding the structure of quantum states. However, it is strictly weaker than local unitary (LU) equivalence [JCWY10, TsGü17, BuJV25, ClPe25b], which encompasses all local quantum operations that leave the overall entanglement unchanged (Figure 5). Recent advances have introduced generalized notions such as R-local complementation to bridge the gap between LC and LU equivalence [ClPe25a], offering a complete picture of graph state transformations, and more efficient procedures to decide LU equivalence [BuJV25, ClPe25b].

Measurement-based quantum computing (MBQC) is a model of quantum computation that uses graph states as the primary resource [RaBr01, RaBB03]. In MBQC, computation proceeds through a sequence of adaptive single-qubit measurements on a pre-constructed entangled state (Figure 6). The logical circuit one wishes to simulate determines the choice of measurement basis and classical processing of outcomes. This model separates the entanglement generation phase from the computational phase, offering advantages in modularity, fault tolerance [RaHG06], potential reductions in quantum circuit depth [BoKa07, BrKa09, LQRS22], and error detection [QASR24].

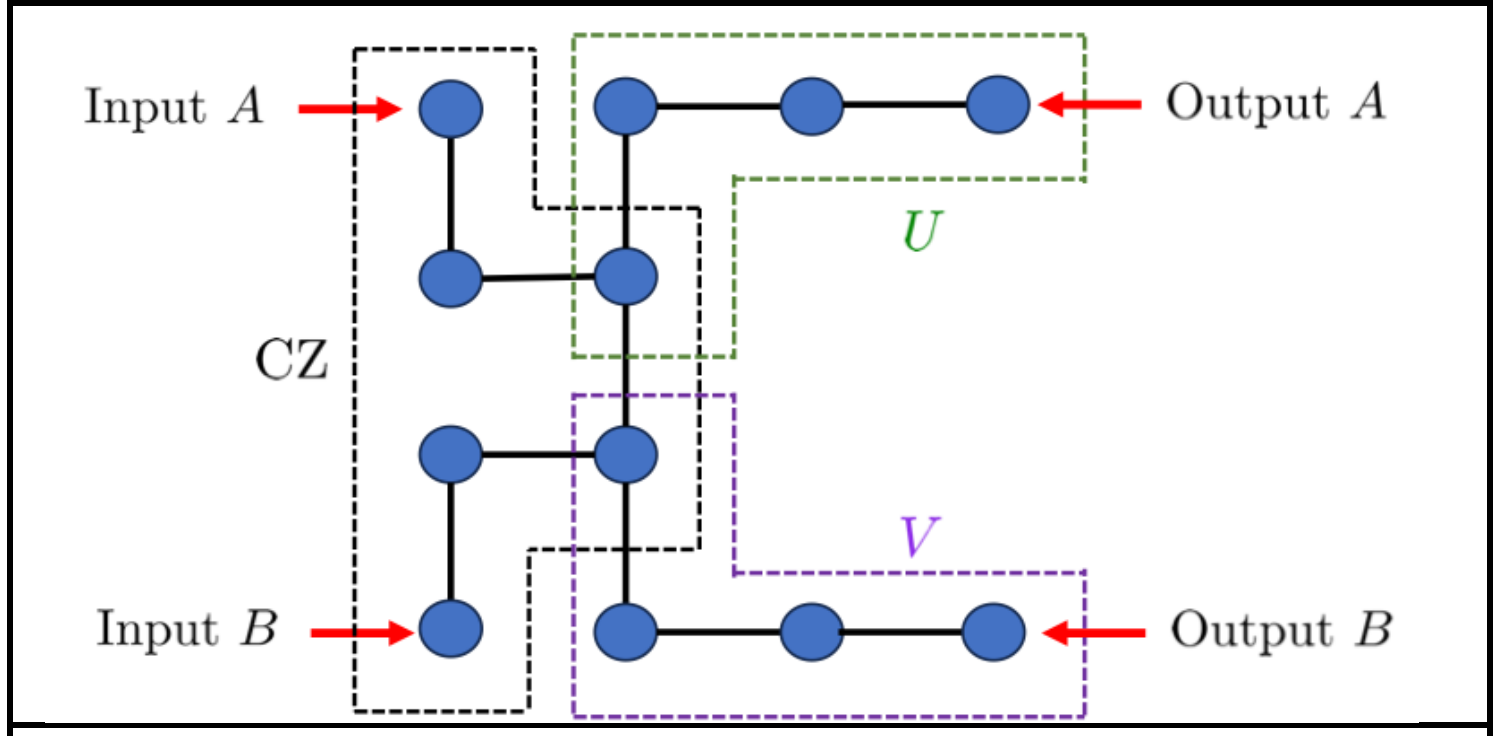

*Figure 6: An example graph state whereupon properly chosen single qubit measurements execute a controlled-Z operation followed by local unitaries, U and V.*

MBQC is equivalent in computational power to the standard quantum circuit model, but it offers a different perspective on how quantum information is processed. For example, certain quantum circuits can be implemented with reduced depth in MBQC by leveraging the parallelism inherent in the measurement process [RaBB03, BrKa09, LQRS22]. Moreover, MBQC has found applications in secure multi-party computation, where graph states enable protocols for secret sharing [MaSa08, JaMP13] and distributed computation with strong privacy guarantees [FiKa17].

## III. Session Findings (Opportunities and Challenges)

The following summarizes findings from discussions that followed presentations in each session of the workshop.

### A. Quantum Graph States and Quantum Information

The session on quantum graph states and quantum information covered a wide range of topics, with a particular focus on the characterization of computational resources. One central theme was the role of graph-theoretic properties in determining the usefulness of quantum states for MBQC [RaBr01, RaBB03].

A key example discussed was the concept of generalized flow (gflow) [BKMP07]: a graph property that is efficiently computable and characterizes those graph states that support deterministic computation, i.e., for which a correction strategy exists. Such strategies are essential to counter the inherent non-determinism of quantum measurements. Generalizations of gflow were also explored [MhPS25], and several open questions related to these extensions were raised.

A particularly compelling research direction is the effort to characterize the resources required for MBQC in terms of graph theory: identifying universal resource states, understanding the types of entanglement that can be leveraged, and determining how these questions translate not only into the language of graph states, but also into the more general framework of hypergraph states [RHBM13], which offer promising advantages [MiMi18, GaGM19].

The session also touched on the crucial issue of fault-tolerant quantum computing, a pervasive challenge in the field. The graph state formalism was highlighted as a powerful tool for the design and implementation of quantum error-correcting codes, and more broadly, for building computation schemes that are resilient to noise.

Additionally, the classical simulation of graph state-based computations was discussed. Structural properties of graphs such as rank-width and tree-width were shown to impact simulability. The roles of vertex-minors and local Clifford equivalence in determining computational power and simulation complexity were also examined.

The session concluded with lively and insightful discussions on the need for new graph-theoretic tools to advance the field of quantum information science, highlighting the deep and growing interplay between quantum computation and combinatorial graph theory.

### B. Advances in Classical Graph Theory and Connections to Quantum Theory

The session on graph theory and networking focused on a variety of topics. These included the tree-width of graphs [RoSe84] (an analogue of the rank-width, which is known to be a criterion for MBQC [VDVB07]), potential applications of entanglement to combinatorial games (e.g., the graph isomorphism game [MaRo20] and Levine's hat problem [AFKK23]), using quantum computing tools to speed up algorithms for combinatorial problems (e.g., faster graph coloring algorithms [FeHC25]), and how qubit topologies affect the performance of certain applications (e.g., quantum routing [DSBC24]).

One interesting line of thought involved the rank-width (see Reference [Oum17] for a definition), and how it can be interpreted from a quantum perspective. The relation between the rank-width and MBQC was discussed [VDVB07,GDHG23], but also more recent work demonstrating tantalizing connections between the rank-width of a graph state with the complexity of creating the state in a lab [KuMY25, DaJe25]. It was established that an important direction lies in extending such results to more realistic settings, where the difficulty of implementing two-qubit gates is not uniform.

Combinatorial algorithms are concerned with optimizing over a set of combinatorial objects. Such algorithms are not only interesting from a theoretical point of view but have also been widely applied in practical settings such as operations research. In certain cases, the (combinatorial) structure of such problems enables techniques from quantum computing. During the session potential applications and tools from quantum were discussed. Of interest here are protocols that move beyond applying Grover's algorithm, and instead exploit the structured nature of the problem.

Combinatorial games also came up during discussion, e.g., the (quantum) graph isomorphism problem. In this problem, two provers share entanglement, and are interrogated by a verifier. The provers aim to convince the verifier that two given graphs G and H are isomorphic. Without entanglement, the two provers can convince the verifier of the fact that G and H are isomorphic if and only if the two graphs are in fact isomorphic. If, however, the prover is not aware that the two verifiers share entanglement, the prover can be 'tricked' into believing that two non-isomorphic graphs are, in fact, isomorphic.

Highlights of the last session included the behavior of the rank-width for k-regular graphs [GDHG23] and modular decompositions of states/qubit topologies (as motivated by experimental constraints). One key point was that optimizing modular decompositions is complicated not only by the variety of metrics (such as application/state creation runtime), but also on the set of allowed operations. As an example, allowing for local operations and classical communication reduces the time to generate a GHZ state from linear to constant time [StCG25]. There was significant interest in whether such statements could be made for general graph states. Furthermore, differences between Hamiltonian- and circuit-based approaches were also highlighted.

## C. Experimental Progress on the Construction and Use of Quantum Graph States

The session on quantum graph states brought together experimentalists and theorists to explore the current landscape and future directions of graph state research across various quantum platforms. The discussion was structured around five guiding questions, focusing on applications, hybrid systems, scalability, theoretical needs, and experimental milestones. The following summarizes key findings.

*Opportunities and emerging directions*: One of the most compelling themes was the potential of hybrid graph states, which involve entangling qubits across different physical platforms—such as atoms and photons, or discrete and continuous variable (CV) systems. These hybrid systems are seen as promising candidates for quantum interconnects and transduction, especially in scenarios with photon-mediated entanglement between remote qubits. Cavity QED systems were identified as a natural platform for exploring such hybrid entanglement, with early experimental demonstrations already underway [CKPS24].

Another exciting opportunity lies in the demonstration of macroscopic entanglement. Cluster states, due to their stabilizer structure and local connectivity, exhibit lifetimes that are independent of system size [HeDB05]—unlike GHZ states, which degrade rapidly with increasing qubit number. This robustness makes cluster states ideal candidates for demonstrating large-scale entanglement, a milestone that would not only validate theoretical predictions but also serve as a steppingstone toward fault-tolerant quantum computing.

Large scale cluster states will likely require generation methods that are intrinsically efficient and scalable to the specific qubit systems, not by sequential quantum gates. For example, interactions among multiple atoms to create a cluster state have been discussed. Similar processes may exist for photonic cluster states as well where a central source of optical nonlinearity directly produces a complex state consisting of many photons.

The discussion also touched on continuous variable systems, particularly Gaussian states, which offer efficient methods for measuring entanglement entropy and simulating exotic entanglement structures, such as those predicted by conformal field theories [PCKW21]. These systems could bridge quantum information science with fundamental physics, including potential links to spacetime curvature.

*Challenges and needs*: Despite these opportunities, several challenges were identified. Scalability remains a central concern, with short-term issues like instrumental noise and decoherence, and longer-term hurdles involving the integration of error correction. Participants emphasized the need for platform-specific error mitigation strategies, especially in the pre-error-correction regime where high-fidelity state preparation is essential.

Benchmarking and certification of graph states also emerged as a critical bottleneck. While stabilizer measurements are straightforward for cluster states, they may not always be optimal. Recent work demonstrated benchmarking of graph states with non-local order parameters using Rydberg atom arrays [QiSc25]. GHZ states, in contrast, require full-body correlation measurements, making them fragile and difficult to certify. The design of entanglement witnesses that are both noise-resilient and task-relevant was highlighted as a key theoretical challenge.

Participants also called for smarter benchmarking techniques and theory-driven error mitigation methods tailored to specific platforms. Questions were raised about the applicability of dynamical decoupling in CV systems and the need for better understanding of how entanglement witnesses relate to the practical utility of a state, such as in MBQC.

The session concluded with a consensus that while the path to scalable, fault-tolerant quantum computing is complex, the interplay between theory and experiment—especially in the context of graph states—offers a fertile ground for innovation. Participants expressed optimism that with the right theoretical tools and experimental ingenuity, many of these challenges can be overcome.

## D. A Graph Perspective on Distributed Quantum Sensing

The session on the utility of entanglement triggered discussion on various topics. One of the focuses is distributed quantum sensing [ZhZh21], where entanglement in the spatial domain and frequency domain can assist the acquiring of information.

Quantum sensors utilize quantum effects, such as squeezing and entanglement, to enhance the sensing of force, radiofrequency (rf) signal, magnetic field, gravitational wave and hypothetical dark matter. Distributed quantum sensors are a versatile sensing paradigm where global features of many local parameters can be extracted with enhanced precision. In many cases of distributed quantum sensing, the form of entanglement necessary in such application can be regarded as generalized graph states, such as GHZ states, multi-mode squeezed states [ZhZS18] and states generated from passing two single-mode squeezed vacuum on a beamsplitter network [ZhZh19]. Here, entanglement can refer to quantum correlations between multiple spatial separated sensors or between multiple frequencies of the same sensor. In addition, the optimal form of entanglement relies on the global feature of interest, e.g., the weights of a linear combination, and therefore may need to be learned during a training process when sensing unknown random signals.

In terms of spatial entanglement, relevant applications include entangled optomechanical sensor arrays [XAPB23] and rf sensor arrays [XLCH20]. One type of entanglement beneficial to the sensor network can be generated from passing a squeezed vacuum state through a linear beamsplitter network. In terms of frequency domain entanglement, an example is the dual-comb spectroscopy protocol enhanced by entangled quantum combs [HLSZ24] where modes with different frequencies are entangled.

*Opportunities and challenges:*

1. Enable more drastic (e.g. exponential scaling with number of sensors) advantages over classical counterparts. Typical entangled sensor network only enjoys a polynomial advantage. Recent works [WBSB24] are progressing towards exponential advantage, however, not yet in practically useful applications.
2. Enable quantum advantages robust against noise. Typical quantum advantage from entanglement quickly decays with loss or noise.
3. Non-Gaussian entanglement is widely unexplored. The efficient description and analyses of such multi-partite entangled states, potentially with graph state language, is unclear, for first approaches see Reference [VJXG24]. Also, it is important to consider non-Gaussian states that can be engineered efficiently with near-term devices.
4. Besides sensor networks, entanglement and squeezing has recently been shown to enhance quantum transduction [ShZh24]. It is open whether multipartite entanglement beyond bipartite can enhance quantum transduction.

## E. Quantum Graph State Generalizations

Graph states are a versatile family of quantum states and are relevant for many applications, from quantum sensing to MBQC and quantum error correction. Still, it is highly relevant to consider potential generalizations of graph states. For this, there are several reasons:

- The Gottesman-Knill theorem states that the generation and measurement of stabilizer states (and hence graph states) can be efficiently simulated (when restricted to Clifford operations and Pauli measurements). Therefore, studying extensions of graph states will help to understand the necessary resources for obtaining the full quantum advantage in computation. In a similar fashion, the model of match-gate circuits has been shown to become universal by adding only simple long-range gates [JoMi08].
- For the theory of graph states, the problem of LU vs. LC equivalence is central [VaDD04,ClPe25a,BuJV25,ClPe25b]. It has been shown that this problem can be understood by considering weighted hypergraph states; this allows us to construct counterexamples to the LU-LC conjecture in a systematic manner [TsGü17]. This demonstrates that generalizations of graph states are frequently the natural picture to study questions related to graph states.

For the generalization of graph states, several directions are promising:

- Weighted graph states: These are generated from product states via two-qubit gates with a variable phase, not a fixed phase. This leads to a continuous family of states, for which still many interesting quantities (like reduced density matrices) can efficiently be computed. Weighted graph states have found application in the analysis of decoherence models [CHDB07, YuGN23].

- Hypergraph (HG) states: These are generated from product states via multi–qubit interactions, but still with fixed phases (multi-qubit controlled phase gates). HG states can still be described by a (albeit non-local) stabilizer group; moreover, many transformations and measurements can be described in a graphical manner [RHBM13, GCSM14]. HG states have turned out to be useful for MBQC [GaGM19, MiMi18] and they lead to large non-classical effects, e.g., by violating Bell inequalities in an extreme manner [GaBG16]. Recently, HG states have been observed in photonic experiments [HLCZ24].
- Locally maximally entanglable (LME) states: Most generally, one can consider multi-qubit interactions, but with a variable phase. Such states have first been considered as LME states [KrKr09], alternatively they can be seen as weighted hypergraph states. For them, optimized gate sequences for their generation exist, moreover, they have been used to analyze decoherence processes [CaBK11] and purification protocols have been designed for them [CKDD13].
- High-dimensional extensions: The previous versions of graph states and their generalizations are initially defined for multi-qubit systems, but one can naturally extend them to higher-dimensional or even continuous-variable systems, see, e.g., References [SRMG17] and [VJXG24]. This is in line with recent advances in the creation of high-dimensional entanglement [ErKZ20].

For these generalizations, there are the following opportunities and challenges:

- HG states and Pauli measurements on them can, in general, not efficiently be simulated. It would be interesting to characterize which families of HG states can still be simulated (e.g., potentially certain uniform HG states, where the edges all have only a fixed cardinality) and which ones are difficult to simulate (for instance, one may characterize "magic states" by determining the stabilizer rank or stabilizer extent).
- Generally, for all generalizations of graph states, it is a central challenge to characterize the advantage of these generalizations for quantum information processing tasks. For instance, one may study their usefulness for quantum sensing by determining their quantum Fisher information. From existing works, one knows already that HG states may have some advantage in metrology, as they are more robust against particle loss, compared to the standard GHZ states [GaBG16].
- Generalized graph states and networks: There are many works on distributing and transforming usual GS in distributed scenarios, these ideas need to be extended to generalizations. This includes novel purification protocols (which cannot rely on the standard stabilizer formalism anymore), as well as protocols to characterize the topology of distributed states (see Reference [WPHD24] for the example of GHZ states).

- A further challenge lies in the experimental generation of HG states and LME states: Existing experiments with photons have low efficiencies, but for Rydberg atoms the implementation of the required multi-qubit gates seems feasible [EBKE23]. From the theoretical side, it is highly relevant to develop systematic and robust implementation schemes for HG states and other generalizations of graph states for the different set-ups. Moreover, efficient verification methods (e.g., based on fidelity estimation) are needed.
- A further central interdisciplinary challenge is the connection of mathematical concepts from the theory of hypergraphs to physical properties of HG states. This includes: (i) The development of a graphical calculus for transformation between HG states. (ii) The study of bipartite / tripartite entanglement properties of HG states in terms of graphical invariants, if the N particles are split into two or three groups. (iii) The analysis of the entanglement dimensionality of HG states in higher dimensions, if the multi-qubit gates act on $d$-dimensional qudits (and not only qubits). This may connect the tensor rank of the quantum state to graphical properties. (iv) The study of SLOCC orbits of generalized graph states.

The general aim of this research program is the following: Currently, graph states (and sometimes more specifically GHZ states) form a test-bed in entanglement theory, in the sense that for them many questions can be solved [GüSe10], simplified [JuMG11, CoTa24] or the analysis of graph states helps to characterize quantum correlations for general states by providing lower bounds on entanglement measures [ElSi12]. If this can be extended to generalized graph states, this may boost the theory of entanglement and quantum information theory.

## IV. Research Gaps and Recommendations

This section collects findings from the workshop that identify critical gaps in current knowledge or technology and propose specific research directions or programmatic initiatives that complement those discussed above.

*Overlap between classical and quantum graph theory:*

- Mathematicians are usually interested in vertex-minors up to isomorphisms. Physicists, on the other hand, are concerned with the physical location of their qubits, especially in a networked setting. Can some of the known theorems on vertex-minors be adapted to such a *labelled* setting, and are there interesting mathematical questions to pursue for the labelled setting? As an example, vertex-minor universal graph states are graph states on $n$ qubits such that any graph state on a subset of $k$ vertices is a vertex-minor [CCMP24]. While randomized constructions achieve the optimal scaling of $n \sim k^2$, explicit constructions only achieve $n{\sim}k^4$; can this be improved? One can also be less demanding and only require that a given list of labelled graphs appear as vertex-minors of some desired resource state. Bounds on $n$ or heuristic constructions of such resource states would be powerful.

- Natural classes of graph states are *vertex-minor closed,* meaning that vertex-minors of graphs in the class are in the class as well. Several natural examples include graphs with bounded (linear) rank-width and circle graphs. Large rank-width graph states are necessary for MBQC (see Reference [VDVB07] for more information). But circle graph states (see Reference [DaHW20] for a definition) are an important class of graphs from a mathematical perspective, where they form a structural role like planar graphs in graph minor theory. In the quantum information community, where they have been mostly studied since they are amenable to tools that non-circle graph states are not (see References [DaHW20, DaHW22], and [BhGo25]). A purely quantum-mechanical interpretation (e.g., circle graph states are exactly those states that are not useful for a certain application) is still missing, however.
- The rank-width has been shown to control the minimum number of two-qubit interactions needed to create a graph state [DaJe25, KuMY25]. It is natural to wonder whether resource requirements for devices with—say—limited qubit connectivity can be related to other width parameters (and their associated hierarchical decompositions). Of particular interest here are devices and graph states that have a hierarchical structure [which would facilitate their (experimental) fault-tolerant preparation] and how measurements and feedback/auxiliary qubits could impact such statements.
- Separately, several discussions were had on further research on combinatorial games, such as the (quantum) graph isomorphism problem [MaRo20]. For example, there exist games closely related to the graph isomorphism, based on relational structures [Ciar24] — are there other games that have a quantum advantage that can be interpreted as a type of multi interactive proof system?
- The combinatorial games studied so far are usually of a more theoretical bent, and it is not directly clear what the practical implications could be. An important consideration that came up during discussions was if and how 'quantum cheating' in combinatorial games can be used to mislead verifiers that assume the provers are classical. For the graph isomorphism game in particular, quantum cheating can be prevented by the verifier by storing classical information and asking repeated questions. Is there a way for the provers to get around this? One natural question here is whether having more than two provers makes it harder for the provers to convince the verifier that (say) two non-isomorphic graphs are isomorphic.
- Several questions regarding entanglement routing came up. Already for simple settings like the star graph the fastest way to perform is unknown [DYGS25]. Natural non-trivial extensions include considering noise/non-uniform interaction strengths by considering weighted interaction graphs.

*Experiments designed to build quantum graph states:*

- Schemes for generating entanglement through intrinsic interactions rather than two-qubit gates are highly needed for generating large-scale cluster states.
- Concerted research efforts combining cluster states across different platforms would be valuable.
- Experiments that truly demonstrate the advantages or applications of cluster states, such as non-local security with photonic cluster states, would be important milestones.
- Looking ahead, several experimental milestones were proposed for the next five years:
  - Demonstrating large-scale photonic cluster states with size-independent lifetimes.
  - Certifying entanglement under various noise models, including inhomogeneous environments.
  - Constructing hybrid graph states that link different qubit types or CV/discrete systems.
  - Exploring holographic entanglement structures in CV systems.
  - Developing robust, task-specific entanglement witnesses.

*Generalized Graph States:*

- Concrete protocols for the creation of generalized graph states for the various platforms (mainly photonic qubits and arrays of Rydberg atoms) are needed. This requires interaction between theory and experiment. In addition, efficient verification and characterization methods for generalized graph states, tailored to the specific platforms are required.
- A central challenge is the connection of the mathematical theory of hypergraphs with physical properties of hypergraph states. Elucidating this will improve various applications of hypergraph states in MBQC and quantum sensing and will shed new light on entanglement as a resource in quantum information processing.

## V. Conclusion

The workshop underscored the growing synergy between classical graph theory and quantum information science, revealing how structural graph properties underpin critical aspects of quantum computation, communication, and sensing. Key findings include the identification of rank-width and vertex-minors as central to understanding quantum resource states, the promise of hybrid and generalized graph states for scalable quantum architectures, and the need for robust experimental and theoretical tools to benchmark and harness entanglement. The discussions emphasized the importance of interdisciplinary collaboration to bridge theoretical insights with experimental capabilities, paving the way for breakthroughs in quantum technologies.

## VI. Acknowledgements

We thank the Air Force Office of Scientific Research (Award No. FA9550-25-1-0059), the Office of Naval Research (Award No. N000142512350), the Army Research Office (Award No. W911NF2510057), and Virginia Tech's College of Science and Department of Physics for support.

# VIII. Appendix: Participant List

| | | |
|---|---|---|
| Rad | Balu | Army Research Office |
| Thomas | Barthel | Duke U. |
| Hans | Briegel | U. of Innsbruck |
| Kenneth | Brown | Duke U. |
| Jinglei | Cheng | U. of Pittsburgh |
| Andrew | Childs | U. of Maryland |
| Eric | Chitambar | UIUC |
| Maria | Chudnovsky | Princeton U. |
| Nathan | Claudet | LORIA |
| Julien | Codsi | Princeton U. |
| Roberto | Diener | Office of Naval Research |
| Sophia | Denker | U. Siegen |
| Asaf | Ferber | UC Irvine |
| Sara | Gamble | Army Research Office |
| Kenneth | Goodenough | U. Mass. Amherst |
| Alexey | Gorshkov | U. of Maryland |
| TR | Govindan | Army Research Office |
| Saikat | Guha | U. of Maryland |
| Otfried | Gühne | U. Siegen |
| Rebekah | Herrman | U. of Tennessee |
| Delaram | Kahrobaei | Queens College, CUNY |
| Alicia | Kollár | U. of Maryland |
| Jianqing | Liu | NC State U. |
| Junyu | Liu | U. of Pittsburgh |
| Yi-Kai | Liu | U. of Maryland |
| Zhenning | Liu | U. of Maryland |
| Hsuan-Hao | Lu | Oak Ridge National Laboratory |
| Rose | McCarty | Georgia Tech |
| Grace | Metcalfe | Air Force Office of Scientific Research |
| Akimasa | Miyake | U. of New Mexico |
| Simon | Perdrix | LORIA |
| Olivier | Pfister | U. of Virginia |
| Vito | Scarola | Virginia Tech |
| Ryan | Scott | Virginia Tech |
| Monika | Schleier-Smith | Stanford U. |
| Meg | Shea | Army Research Office |
| Neil | Sinclair | Harvard U. |
| Shuo | Sun | U. of Colorado |
| Mario | Szegedy | Rutgers U. |
| Anna Lina | Vandré | U. Siegen |
| Don | Wagner | Air Force Office of Scientific Research |
| Tzu-Chieh | Wei | Stony Brook U. |
| David | Weiss | Pennsylvania State U. |
| Mark | Wilde | Cornell U. |
| Moe | Win | MIT |
| Kerolos | Yousef | Harvard U. |
| Marat | Valiev | Department of Energy |
| Yanbao | Zhang | Oak Ridge National Laboratory |
| Yichi | Zhang | U. of Pennsylvania |
| Quntao | Zhuang | U. of Southern California |